\documentclass[11pt,twoside]{article}

\usepackage{asp2006}
\usepackage{epsf}
\usepackage{epsfig}
\usepackage{lscape}
\usepackage{natbib}

\begin{document}
\title{Binary sdB Stars with Massive Compact Companions}
\author{S. Geier$^1$, C. Karl$^1$, H. Edelmann$^2$, U. Heber$^1$, and R. Napiwotzki$^3$}   
\affil{$^1$ Dr.--Remeis--Sternwarte, Institute for Astronomy, University Erlangen-Nuremberg, Sternwartstr. 7, 96049 Bamberg, Germany}
\affil{$^2$ McDonald Observatory, University of Texas at Austin,
     1 University Station, C1402, Austin, TX 78712-0259, USA}
\affil{$^3$ Centre of Astrophysics Research, University of Hertfordshire, College Lane, Hatfield AL10 9AB, UK}

\begin{abstract} 
The masses of compact objects like white dwarfs, neutron stars and
black holes are fundamental to astrophysics, but very difficult to
measure. We present the results of an analysis of subluminous B (sdB)
stars in close binary systems with unseen compact companions to derive
their masses and clarify their nature. Radial velocity curves were
obtained from time resolved spectroscopy. The atmospheric parameters
were determined in a quantitative spectral analysis. Based on high
resolution spectra we were able to measure the projected rotational
velocity of the stars with high accuracy. In the distribution of
projected rotational velocities signs of tidal locking with the
companions are visible. By detecting ellipsoidal variations in the
lightcurve of an sdB binary we were able to show that subdwarf
binaries with orbital periods up to $0.6\,{\rm d}$ are most likely
synchronized. In this case, the inclination angles and companion masses of the
binaries can be tightly constrained. Five invisible companions have
masses that are compatible with that of normal white dwarfs or late
type main sequence stars. However, four sdBs have compact
companions massive enough to be heavy white dwarfs ($>1\,{\rm M_\odot}$), neutron stars
or even black holes. Such a high fraction of massive compact
companions is not expected from current models of binary evolution.
\end{abstract}

\section{Introduction}  

The mass of a star is its most fundamental property. However, a
direct measurement is possible in a small number of binary stars
only. White dwarfs, neutron stars and stellar mass black holes are the
fibal products of stellar evolution. In most binaries such faint, compact
objects are outshone by their bright companions and therefore their
orbital motion cannot be measured. Only lower limits
to the companion mass can be derived. With the analysis method shown
here, these limitations can be partly overcome.

Subluminous B stars are considered to be helium core burning stars
with very thin hydrogen envelopes and masses around $0.5\,{\rm
  M_\odot}$.  
Different formation channels have been discussed. As it
turned out, a large fraction of the sdB stars are members of short
period binaries \citep{t19_maxted}. For these systems common-envelope
ejection is the most probable formation channel \citep{t19_han}. In this
scenario two main-sequence stars of different masses evolve in a
binary system. The heavier one will first reach the red giant phase
and fill its Roche lobe. If the mass transfer to the companion is
dynamically unstable, a common envelope is formed. Due to friction the
two stellar cores loose orbital energy, which is deposited within the
envelope and leads to a shortening of the binary period. Eventually
the common envelope is ejected and a close binary system is formed,
which contains a helium core burning sdB and a main sequence
companion. When the latter reaches the red-giant branch, another
common envelope phase is possible and can lead to a close binary with
a white-dwarf companion and an sdB star. All known companions of sdB stars in
such systems are white dwarfs or late-type main-sequence stars. If
massive stars are involved, the primary may evolve into a neutron star
(NS) or a black hole (BH) rather than a white dwarf. Since massive
stars are very rare, only few sdB+NS or sdB+BH systems are expected to
be found.

Since the spectra of most binary sdBs are single-lined,
 they reveal no information about the orbital motion of their
 companions, and 
 thus only their mass functions can be calculated.

 \begin{equation}
 \label{t19:equation-mass-function}
 f_{\rm m} = \frac{M_{\rm comp}^3 \sin^3i}{(M_{\rm comp} +
   M_{\rm sdB})^2} = \frac{P K^3}{2 \pi G} .
 \end{equation}

Although the RV semi-amplitude $K$ and the period $P$ are determined
 by the RV curve, two of $M_{\rm sdB}$, $M_{\rm comp}$ and $\sin{i}$ remain
 free parameters.
Binary population synthesis models \citep{t19_han} indicate a possible
mass range of $M_{\rm sdB}$\,=\,0.30$-$0.48\,M$_{\rm \odot}$ for sdBs
in binaries, which underwent the common envelope ejection channel. The
mass distribution shows a sharp peak at about $0.46\,{\rm M_\odot}$. 
This theoretical value can be backed up by observations
(Edelmann these proceedings; Vuckovic et al. these proceedings) as
well as asteroseismology (e.g. Charpinet et al. these proceedings).

For close binary systems, the components' stellar rotational
velocities are considered to be tidally locked to their orbital
motions, which means that the orbital period of the system equals the
rotational period of the companions. If the companions are
synchronized in this way the rotational velocity $v_{\rm rot}$ can be
calculated.

\begin{equation}
v_{\rm rot} = \frac{2 \pi R_{\rm sdB}}{P} .
\end{equation}

The stellar radius $R$ is given by the mass--radius--relation.

\begin{equation}
R = \sqrt{\frac{M_{\rm sdB}G}{g}}
\end{equation}

The measurement of the projected rotational velocities 
 $v_{\rm rot}\,\sin\,i$ 
therefore allows us to constrain the systems' inclination angles $i$.
With $M_{\rm sdB}$ as free parameter the mass function can be solved
and the inclination angle as well as the companion mass can be
derived. Because $\sin{i} \leq 1$, a lower limit for the sdB mass is
given.  To constrain the system parameters in this way it is necessary
to measure $K$, $P$, $\log{g}$ and $v_{\rm rot}\sin{i}$ with high
accuracy.

\section{Observations and Radial Velocity Curves}

Ten stars were observed at least twice 
 with the high-resolution spectrograph UVES at the ESO\,VLT.
Additional observations were made 
 at the ESO\,NTT (equipped with EMMI), the Calar Alto Observatory 3.5-m telescope
 (TWIN) and the 4-m WHT (ISIS) on La Palma.
Two of the stars (PG\,1232$-$136, TONS\,183) were observed with the
high resolution FEROS instrument at the 2.2\,m ESO telescope at La
Silla.  The radial velocities (RV) were measured by fitting a set of
mathematical functions (Gaussians, Lorentzians and polynomials) to the
hydrogen Balmer lines. Sine curves were fitted to the RV data points
using a \(\chi^{2}\) minimising method (singular-value decomposition) 
and the temporal power spectrum was generated to obtain a best-fit
 period (Table~1).

\section{Atmospheric parameters and projected rotational velocities}

The spectra were corrected for the measured RV and
 co-added. Atmospheric parameters were determined by fitting
 simultaneously each observed hydrogen and helium line with a grid of
 metal-line blanketed LTE model spectra. Partial results are listed in
 Table~1.

\begin{table} 
\label{t19:param}
\caption{Surface gravities, orbital periods,
radial velocity semi-amplitudes and projected rotational velocities of
the visible components.  The typical error margin for 
$\log g$ is 0.05\,dex.  $\dag$ Orbital
parameters of this system taken from Geier et al. (these proceedings).
$\ddag$ Orbital parameters of this system taken from
\citet{t19_napiwotzki2}.}
\smallskip
\begin{center}
\begin{tabular}{lccccc}
\tableline
\noalign{\smallskip}
System & $\log{g}$ & $P$ & $K$ & $v_{\rm rot}\,\sin\,i$ \\
       & [${\rm cm^{2}s^{-1}}$] & [d] & [${\rm km\,s^{-1}}$] & [${\rm km\,s^{-1}}$] \\ 
\noalign{\smallskip}
\tableline
\noalign{\smallskip}
HE\,0532$-$4503 & 5.32 & 0.26560 $\pm$ 0.00010 & 101.5 $\pm$ 0.2 & 11.1 $\pm$ 0.6 \\
PG\,1232$-$136 & 5.62 & 0.36300 $\pm$ 0.00030 & 129.6 $\pm$ 0.04 & 6.2 $\pm$ 0.8 \\
WD\,0107$-$342$\dag$ & 5.32 & 0.37500 $\pm$ 0.05000 & 127.0 $\pm$ 2.0 & 20.4 $\pm$ 0.9 \\
HE\,0929$-$0424 & 5.71 & 0.44000 $\pm$ 0.00020 & 114.3 $\pm$ 1.4 & 7.1 $\pm$ 1.0 \\
HE\,0230$-$4323 & 5.60 & 0.44300 $\pm$ 0.00050 & 64.1 $\pm$ 1.5 & 12.7 $\pm$ 0.7 \\
TONS\,183 & 5.20 & 0.82770 $\pm$ 0.00020 & 84.8 $\pm$ 1.0 & 6.7 $\pm$ 0.7 \\
HE\,2135$-$3749 & 5.84 & 0.92400 $\pm$ 0.00030 & 90.5 $\pm$ 0.6 & 6.9 $\pm$ 0.5 \\
HE\,1421$-$1206 & 5.55 & 1.18800 $\pm$ 0.00100 & 55.5 $\pm$ 2.0 & 6.7 $\pm$ 1.1 \\
HE\,1047$-$0436$\ddag$ & 5.66 & 1.21325 $\pm$ 0.00001 & 94.0 $\pm$ 3.0 & 6.2 $\pm$ 0.6 \\
HE\,2150$-$0238 & 5.83 & 1.32090 $\pm$ 0.00500 & 96.3 $\pm$ 1.4 & 8.3 $\pm$ 1.3 \\
HE\,1448$-$0510 & 5.59 & 7.15880 $\pm$ 0.01300 & 53.7 $\pm$ 1.1 & 6.7 $\pm$ 2.5 \\
WD\,0048$-$202  & 5.50 & 7.44360 $\pm$ 0.01500 & 47.9 $\pm$ 0.4 & 7.2 $\pm$ 1.3 \\
\hline
\\
\end{tabular}
\end{center}
\end{table}

In order to derive $v_{\rm rot}\,\sin\,i$, we compared the observed
 spectra with rotationally broadened, synthetic line profiles.  The
 latter were computed using the LINFOR program
 \citep{t19_lemke}. Since sharp metal lines are much more sensitive to
 rotational broadening than Balmer or helium lines, all visible metal
 lines were included. A simultaneous fit of elemental abundance and
 projected rotational velocity was performed separately for every
 identified line using the FITSB2 routine \citep{t19_napiwotzki}. The mean
 value and the standard deviation were calculated from all
 measurements.  Seeing induced variations in the instrumental profile
 and the noise level were the dominant error sources. Information on
 the actual seeing conditions for every exposure have been extracted
 from the ESO seeing monitor archive.
All other possible sources of systematic errors turned out to be negligible.

\begin{figure}[t!]
 
 \plottwo{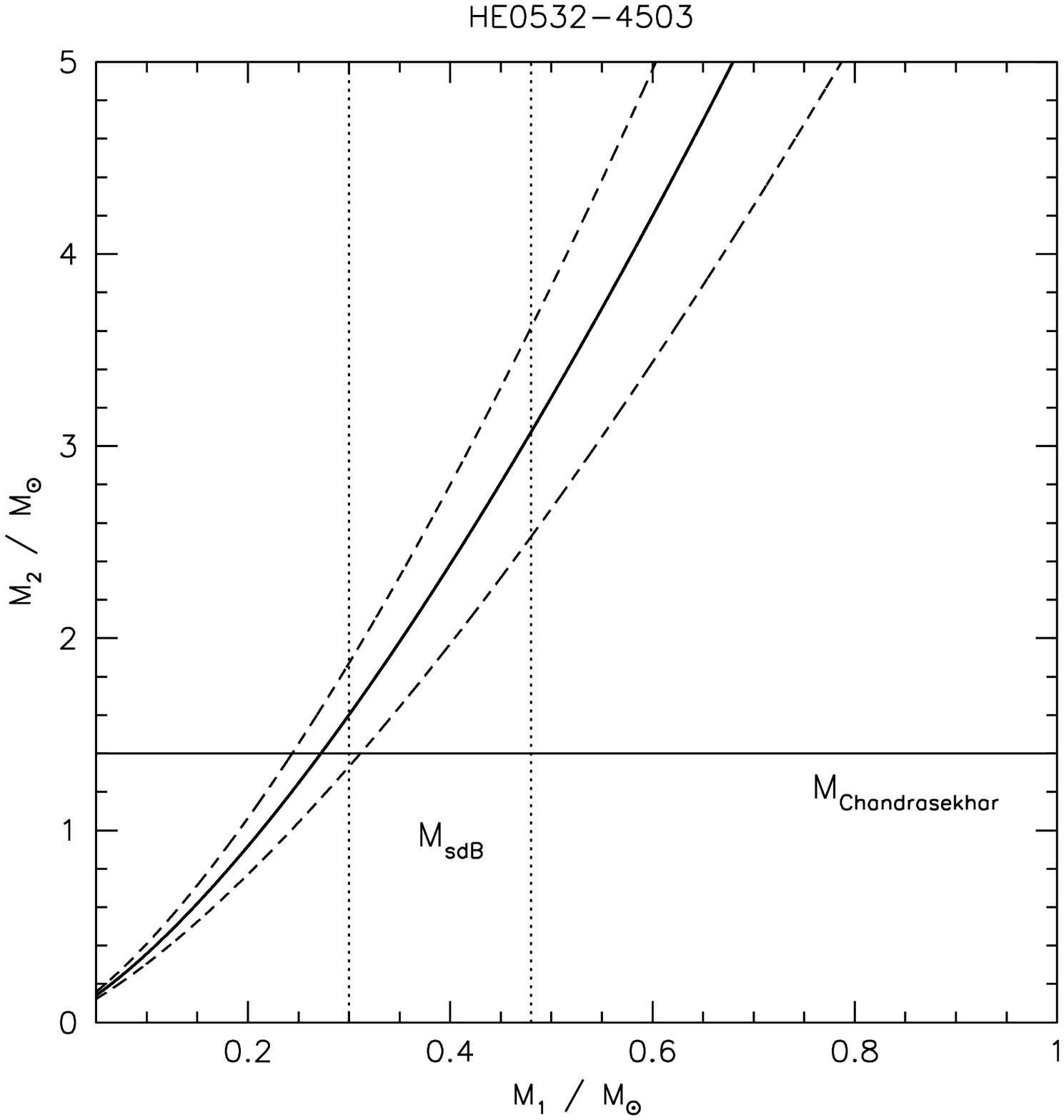}{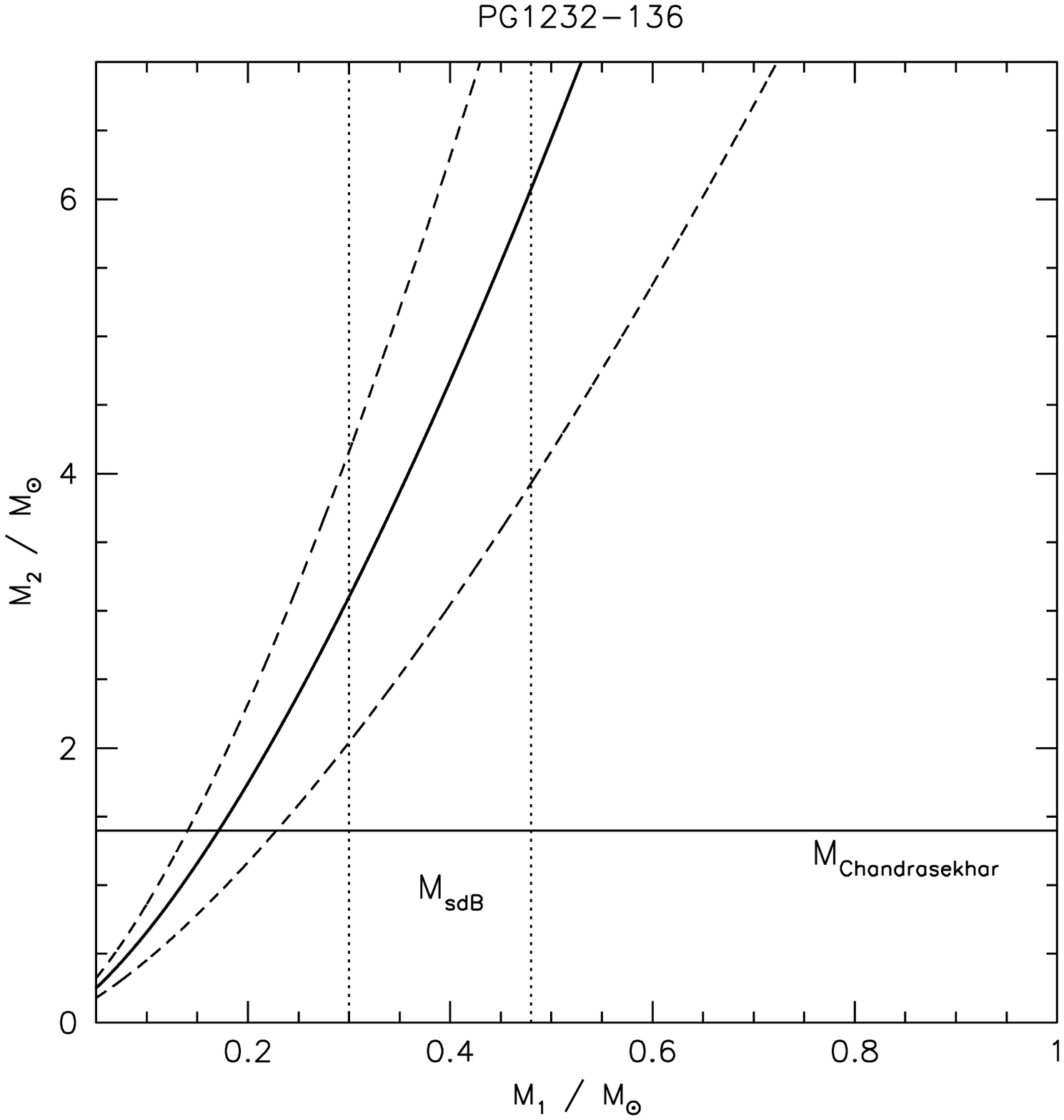}
 \caption{Companion mass as a function of primary (sdB) mass of the
 binaries HE\,0532$-$4503 (left hand panel) and PG\,1232$-$136 (right
 hand panel). The horizontal line marks the Chandrasekhar limit. The
 dotted vertical lines mark the theoretical sdB mass range for the
 common envelope ejection channel \citep{t19_han}.}

 \label{t19:lock1}
\end{figure}

\section{Nature of the Unseen Companions}

Knowing $P$, $K$, $\log\,g$ and $v_{\rm rot}\sin{i}$ we can calculate the
mass of the unseen companion from Eqns. 1$-$3 for any given primary
mass (see Fig.~\ref{t19:lock1}). We adopt the mass range from \citet{t19_han},
marked by the dotted lines in Fig.~\ref{t19:lock1}.  There are no spectral
signatures of companions visible. Main sequence stars with masses
higher than $0.45\,{\rm M_\odot}$ could therefore be excluded because
they would contribute to the total flux and could therefore be
identified from the spectral energy distribution and/or indicative
spectral features. The possible companion masses can be seen in
Table~2. Four of the analysed systems have companion masses, which are
compatible with either typical white dwarfs (WD) or late main-sequence
stars (late MS). The companion of WD\,0107$-$0342 is a massive white
dwarf. Since the total mass of the binary may exceed the Chandrasekhar
limit, the system is a good candidate for a double degenerate SN Ia
progenitor with subdwarf primary, only the second one after
KPD\,1930$+$2752 \citep{t19_geier}.

\begin{figure}[t!]
 \plottwo{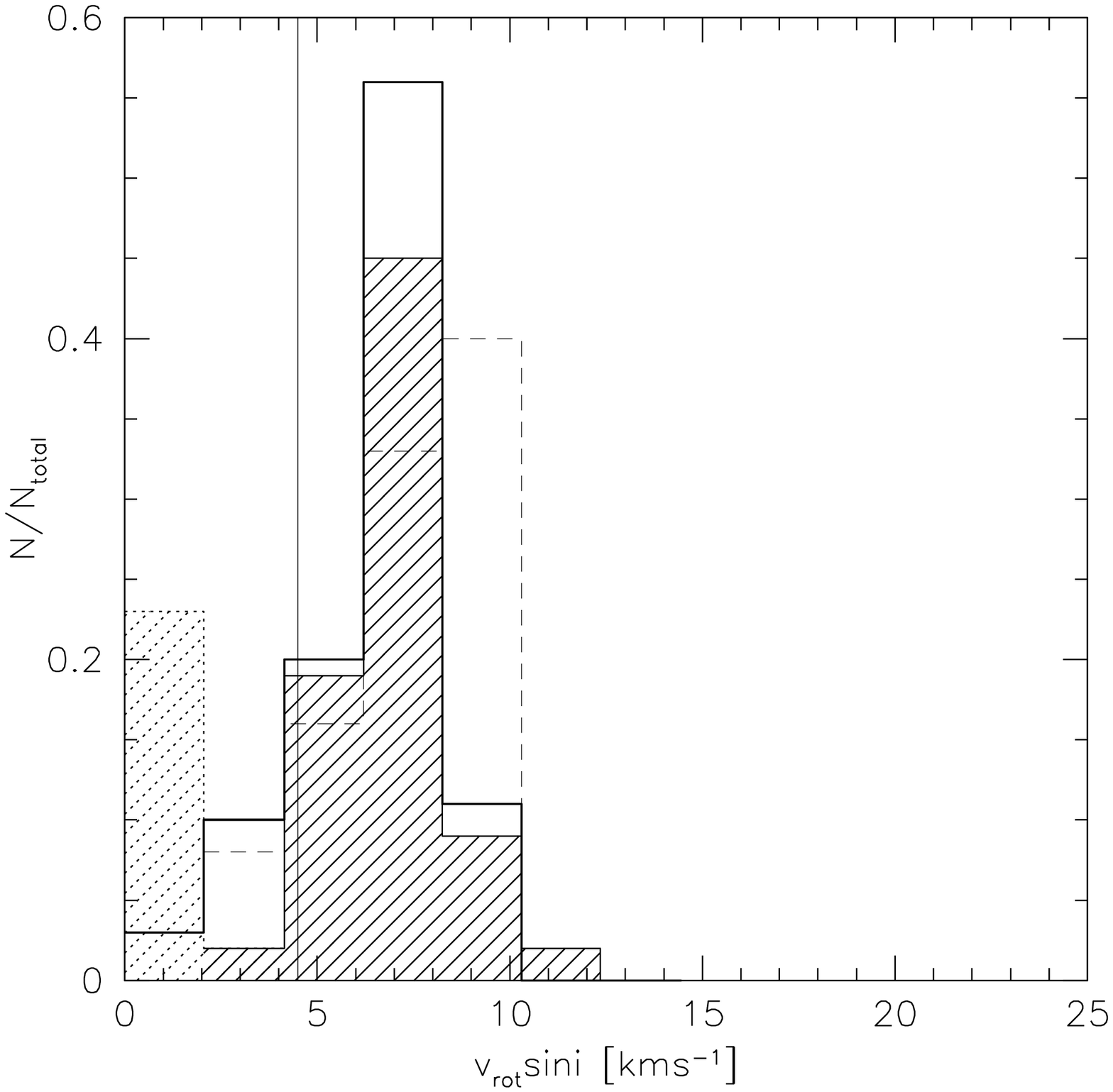}{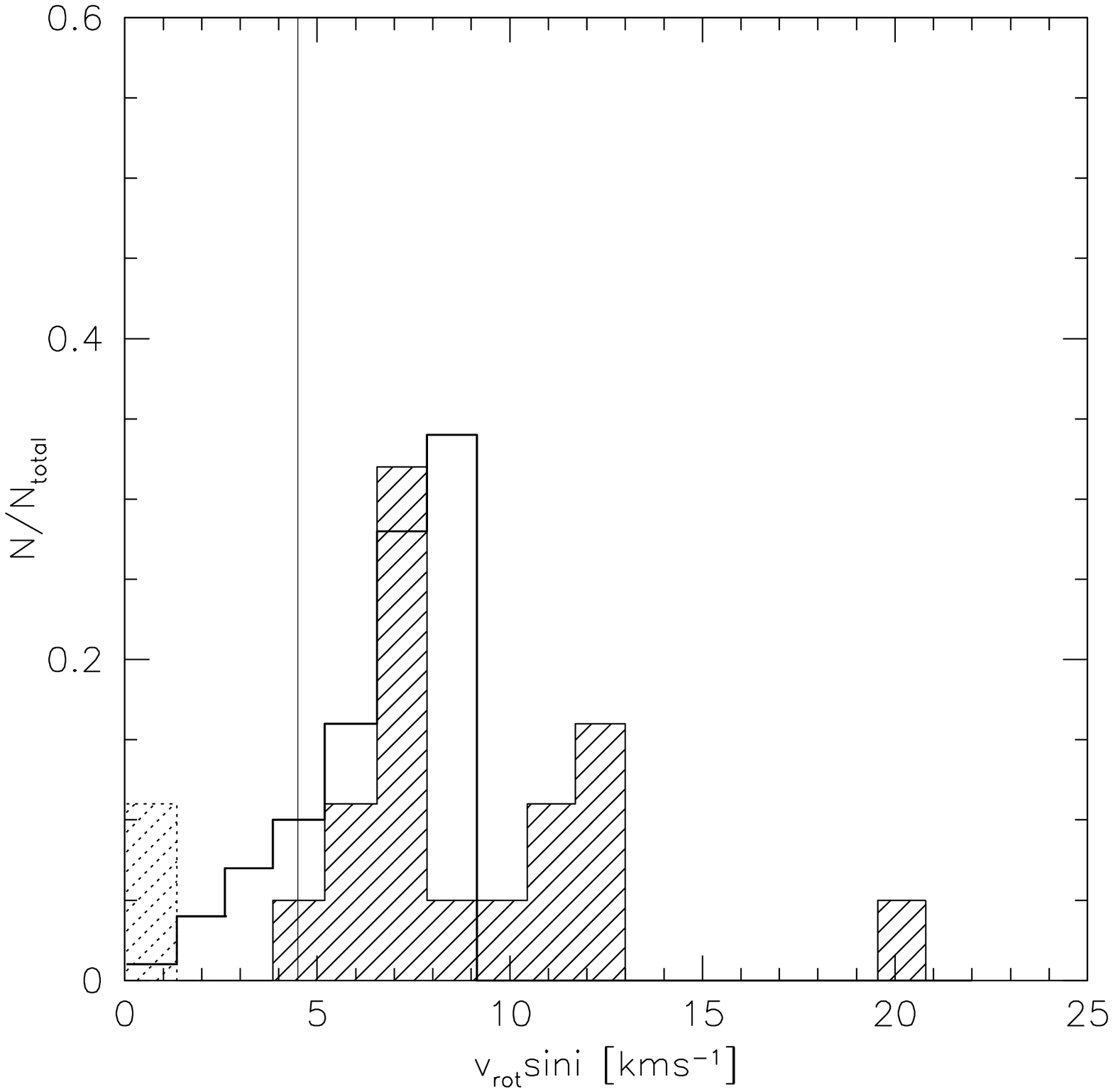}
 \caption{{\it Left panel:} The measured $v_{\rm rot}\sin{i}$ of 49 
 single sdBs is plotted against relative fraction of stars as shaded 
 histogram. The solid blank histogram marks the expected uniform distribution 
 of $v_{\rm rot}\sin{i}$ under the assumption of randomly oriented 
 polar axes and uniform rotational velocity $v_{\rm rot}=8.3\,{\rm kms^{-1}}$ 
 for all stars. The dashed blank histogram marks a uniform distribution for $v_{\rm rot}=9.0\,{\rm kms^{-1}}$ as comparison. 
 All sdBs with $v_{\rm rot}\sin{i}$ below the detection limit 
 (vertical line) are stacked into the first bin (dotted shaded histogram). 
 The width of the bins is given by the average $v_{\rm rot}\sin{i}$ uncertainty 
 of the sample. {\it Right panel:}
The measured $v_{\rm rot}\sin{i}$ of 19 RV variable sdBs is
plotted against relative fraction of stars as shaded histogram. The solid blank histogram marks the uniform distribution derived from the single star sample. The width of the bins is again given by the average $v_{\rm rot}\sin{i}$ uncertainty of the sample, which is lower for the RV variable sample due to better quality of the data.}
 \label{t19:vrotdistrib}
\end{figure}

The very similar HE\,0929$-$0424 and TONS\,183 have to have quite
massive companions. Even in the less likely case that their sdB
primaries are of low mass ($\approx 0.3\,{\rm M_\odot}$) the companions
would be heavy white dwarfs. At the most probable sdB mass, however,
their companions would exceed the Chandrasekhar mass limit. There are
only two kinds of objects known with such high masses and such low
luminosities -- neutron stars and stellar mass black holes. 
The two systems HE\,0532$-$4503 and PG\,1232$-$136 have even
higher companion masses, which would exceed the Chandrasekhar limit
for any mass of a core helium-burning subdwarf star (see
Fig.~\ref{t19:lock1}). The three long-period binaries HE\,1448$-$0510,
HE\,2150$-$0238 and WD\,0048$-$202 could not be solved with the
described method. These systems cannot be synchronized (see Sect. 5).

Binaries hosting a neutron star or a black hole are a very rare class
of objects. From about 50 analysed sdB binaries in our samples, we
found two, possibly four candidate systems. This fraction of 4--8\% is
much too high to be compatible with any binary evolution model known
so far (Podsiadlowski priv. comm.).

Our results depend strongly on the projected rotational velocity. The
unexpectedly high masses could only be reduced if the $v_{\rm
rot}\sin{i}$ were underestimated. As described above we quantified all
possible systematic effects and the overall results were very
consistent. But even if there would be unaccounted systematic effects
(e.g. short period pulsations), they would always cause extra
broadening of the lines. The measured broadening is then due to
rotation plus the unaccounted effects, which means the deduced
rotational broadening would be overestimated. Potential systematic
effects yet unaccounted for would therefore lead to even higher
companion masses.

\section{Orbital Synchronization of sdB Binaries}

Our method rests on the assumption of orbital synchronization. In
Fig.~\ref{t19:vrotdistrib} we compare the $v_{\rm
rot}\sin{i}$-distribution of close sdB binaries to the distribution of
the single stars. A large fraction of binary sdBs exceeds the maximum
$v_{\rm rot}=8.3\,{\rm kms^{-1}}$, derived for the single sdB stars,
significantly. The most likely reason for this is tidal interaction
with the companion.

The question of which mechanisms are responsible for the synchronization
of such stars is not yet settled. Subluminous B stars have convective
cores and radiative envelopes. The physical mechanisms for tidal
dissipation in such stars are under debate.  Different theoretical
concepts \citep{t19_zahn, t19_tassoul} predict that the synchronization
timescale $t_{\rm sync}$ depends strongly on the orbital period but
differs in absolute values.  Tidal locking can only be assumed if the
synchronization time scales are significantly lower than the average
lifetime on the EHB of $t_{\rm EHB}\approx10^8\,{\rm yr}$. Using the
formalism of \citet{t19_zahn} $t_{\rm sync}$ would exceed t$_{\rm EHB}$
for periods longer than $P_{\rm Zahn}^{\rm lim}\approx0.4\,{\rm d}$,
whereas this would be the case at $P_{\rm Tassoul}^{\rm
lim}\approx2.0\,{\rm d}$ if we apply the theory of \citet{t19_tassoul}.

But as long as the question of tidal synchronization is not settled,
all timescales have to be taken with caution \citep{t19_zahn2}. Detailed
calculations, which take the internal structure of sdBs into account,
are not available and urgently needed. Observational constraints are
necessary to guide the development of more sophisticated models.

\begin{figure}[t!]
  \plotone{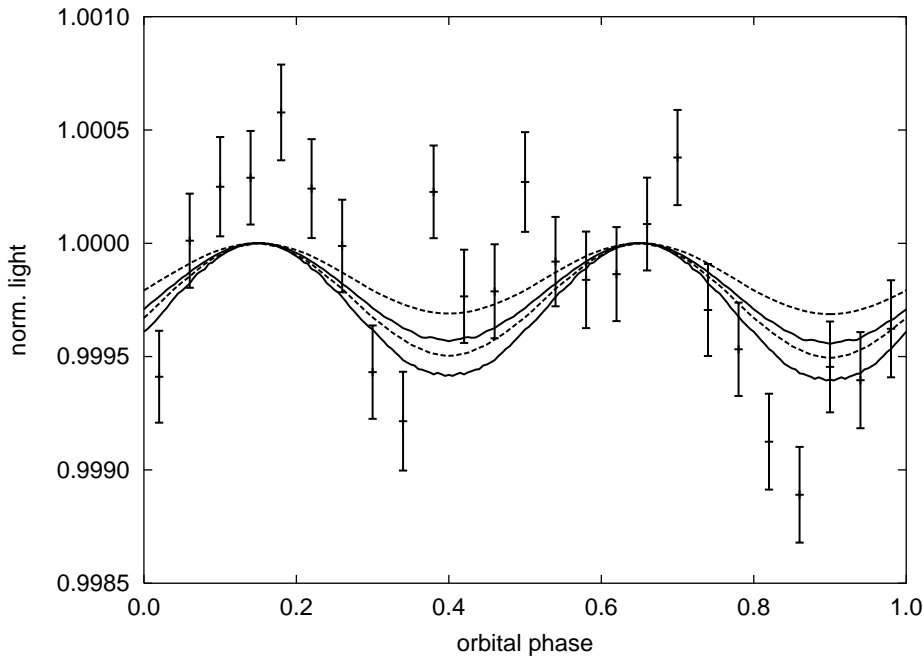}
 \caption{Lightcurve of PG\,0101$+$039 with superimposed models
\citep{t19_geier2}.  The two solid curves confine the best fit models with
parameters and associated uncertainties derived from spectroscopy
under the assumption of orbital synchronization and $M_{\rm
sdB}=0.3\,M_{\odot}$. The two dotted curves confine models for $M_{\rm
sdB}=0.7\,M_{\odot}$.}
 \label{t19:feige11}
\end{figure}

Up to now such observational constraints are rare. Two short period
($\approx 0.1\,{\rm d}$) sdB binaries with white-dwarf companions
show light variations due to their ellipsoidal deformation
\citep[][]{t19_orosz, t19_geier} and are therefore most likely
synchronized. \citet*{t19_randall} reported the detection of shallow
ellipsoidal variations in the longer period ($\approx 0.6\,{\rm d}$)
sdB+WD binary PG\,0101$+$039. \citet{t19_geier2} verified that the light
variations are due to ellipsoidal deformation and that tidal
synchronization is very likely established for PG\,0101$+$039 (see
Fig.~\ref{t19:feige11}). We conclude that this assumption should hold for
all sdB binaries with orbital periods of less than half a day. Last
but not least, van Grootel et al. (these proceedings) carried out an
asteroseismological analysis of the pulsating sdB in the binary Feige
48 (sdB+WD, $P=0.376\,{\rm d}$) and found it to be synchronized.

In Table~2 we show which binaries fulfill the empirical or
theoretical criteria for synchronization. It has to be pointed out
that three of our candidate systems with massive compact companions
are consistent with all of them.

\section{Conclusion}

Out of 12 analysed sdB binaries, five have companion masses compatible
with white dwarfs of typical mass or late-type main-sequence
stars. The properties of these systems are in full agreement with
binary population-synthesis simulations. Four binaries have
surprisingly high companion masses, which leads to the conclusion,
that the companions have to be white dwarfs of unusually high mass or
even neutron stars or black holes. This high fraction cannot be
explained with current evolutionary calculations. As our analysis
assumes synchronization, a better understanding of this process for sdB
stars is urgently needed. A larger sample of sdB binaries has to be
studied with our method to search for systematic trends and improve
statistics.

The presence of such a high fraction of heavy binaries in our samples
raises several questions. Is the formation and evolution of sdB stars
linked to that of heavy compact objects like neutron stars or black
holes? Can sdB stars be used as tracers to find more of these exotic
objects? Is there a hidden population of these objects present in our
galaxy?

\begin{table}[ht!] 
\label{comp}
\caption{Inclination angles, rotational velocities, companion masses
and possible nature of the unseen companions. The lower companion mass
corresponds to an sdB of $0.3\,{\rm M_{\odot}}$, the higher limit to an sdB
of $0.48\,{\rm M_{\odot}}$. Binaries up to orbital periods of $1.3\,{\rm d}$
can be solved consistently under the assumption of
synchronization. Empirical and theoretical indications for
synchronization are given for individual stars. }
\smallskip
\begin{center}
\begin{tabular}{lcccccc}

\tableline
\noalign{\smallskip}

System &  $i$ & $v_{\rm rot}$ & $M_{\rm comp}$ & Companion\\
       & [deg] & [${\rm km\,s^{-1}}$] & [${\rm M_\odot}$] \\ 
\noalign{\smallskip}
\tableline
\noalign{\smallskip}
HE\,0532$-$4503$^{1,2,3,4}$ & 13 $-$ 17 & 47 & 1.40 $-$ 3.60 & \bf NS/BH\\
PG\,1232$-$136$^{1,2,3,4}$ & 14 $-$ 19 & 25 & 2.00 $-$ 7.00 & \bf NS/BH \\
WD\,0107$-$342$^{1,2,3,4}$ & 37 $-$ 50 & 33 & 0.48 $-$ 0.87 & WD \\
HE\,0929$-$0424$^{2,4}$ & 23 $-$ 29 & 18 & 0.60 $-$ 2.40 & \bf WD/NS/BH \\
HE\,0230$-$4323$^{2,4}$ & 38 $-$ 50 & 21 & 0.18 $-$ 0.35 & WD/late MS\\
TONS\,183$^{4}$ & 22 $-$ 29 & 18 & 0.60 $-$ 2.40 & \bf WD/NS/BH \\
HE\,2135$-$3749$^{4}$ & 66 $-$ 90 & 8 & 0.35 $-$ 0.45 & WD/late MS\\
HE\,1421$-$1206$^{4}$ & 56 $-$ 90 & 8 & 0.15 $-$ 0.30 & WD/late MS\\
HE\,1047$-$0436$^{4}$ & 62 $-$ 90 & 7 & 0.35 $-$ 0.60 & WD/late MS\\
HE\,2150$-$0238$^{4}$ & -- & -- & -- & no solution \\
HE\,1448$-$0510 & -- & -- & -- & no solution \\
WD\,0048$-$202  & -- & -- & -- & no solution \\
\hline\\
\end{tabular}

\end{center}

{\small ${^1}$Empirical indications for synchronization could be found in subdwarf binaries with longer periods by asteroseismological modelling.
 
${^2}$Empirical indications for synchronization could be found in subdwarf binaries with longer periods by detection of ellipsoidal deformation. 

${^3}$Theoretical synchronization timescales are shorter than the average EHB lifetime according to the theory of \citet{t19_zahn}.

${^4}$Theoretical synchronization timescales are shorter than the average EHB lifetime according to the theory of \citet{t19_tassoul}}.

\end{table}

\end{document}